\documentclass[twocolumn,aps,prl,superscriptaddress]{revtex4-2}
\usepackage{natbib}
\usepackage[T1]{fontenc}
\usepackage[latin9]{inputenc}
\usepackage{babel}
\usepackage{float}
\usepackage{physics}
\usepackage{amsmath}
\usepackage{amssymb}
\usepackage{amsfonts}
\usepackage{xcolor}
\usepackage{bbm}
\usepackage[caption=false,font=normalsize,labelfont=sf,textfont=sf]{subfig} 
\usepackage{graphicx,epstopdf}
\usepackage{url}
\usepackage{textcase}
\usepackage{bm}
\usepackage[unicode=true]{hyperref}
\usepackage{listings}
\makeatletter

\hypersetup{colorlinks=true, linkcolor=blue, filecolor=blue, urlcolor=cyan,}
\makeatletter
\newcommand*\bigcdot{\mathpalette\bigcdot@{.5}}
\newcommand*\bigcdot@[2]{\mathbin{\vcenter{\hbox{\scalebox{#2}{$\m@th#1\bullet$}}}}}
\makeatother

\begin{document}
\title{Non-Markovian anti-parity-time symmetric systems: theory and experiment}
\author{Andrew Wilkey}
\affiliation{Department of Physics, Indiana University Purdue University Indianapolis (IUPUI), Indianapolis, Indiana 46202, USA}
\author{Joseph Suelzer}
\affiliation{Department of Physics, Indiana University Purdue University Indianapolis (IUPUI), Indianapolis, Indiana 46202, USA}
\affiliation{Air Force Research Laboratory, 2241 Avionics Circle, Wright-Patterson AFB, Ohio 45433, USA }
\author{Yogesh N. Joglekar}
\email{yojoglek@iupui.edu}
\affiliation{Department of Physics, Indiana University Purdue University Indianapolis (IUPUI), Indianapolis, Indiana 46202, USA}  
\author{Gautam Vemuri}
\email{gvemuri@iupui.edu}
\affiliation{Department of Physics, Indiana University Purdue University Indianapolis (IUPUI), Indianapolis, Indiana 46202, USA}
 
\begin{abstract}
Open systems with anti parity-time (anti $\mathcal{PT}$-) or $\mathcal{PT}$ symmetry exhibit a rich phenomenology absent in their Hermitian counterparts. To date all model systems and their diverse realizations across classical and quantum platforms have been local in time, i.e. Markovian.  Here we propose a non-Markovian system with anti-$\mathcal{PT}$-symmetry where a single time-delay encodes the memory, and experimentally demonstrate its consequences with two time-delay coupled semiconductor lasers. A transcendental characteristic equation with infinitely many eigenvalue pairs sets our model apart. We show that a sequence of amplifying-to-decaying dominant mode transitions is induced by the time delay in our minimal model. The signatures of these transitions quantitatively match results obtained from four, coupled, nonlinear rate equations for laser dynamics, and are experimentally observed as constant-width sideband oscillations in the laser intensity profiles. Our work introduces a new paradigm of non-Hermitian systems with memory, paves the way for their realization in classical systems, and may apply to time-delayed feedback-control for quantum systems. 
\end{abstract}

\maketitle

\noindent{\bf Introduction.} Since the seminal work of Bender and co-workers~\cite{Bender1998,Bender2001},  the field of non-Hermitian Hamiltonians with parity-time ($\mathcal{PT}$) symmetry has diversified and matured over the past two decades~\cite{Feng2017,ElGanainy2018,Miri2019,Sahin2019}. $\mathcal{PT}$-symmetric Hamiltonians represent open classical systems with balanced gain and loss~\cite{Joglekar2013}, and have been experimentally realized in diverse platforms comprising optics~\cite{Guo2009,Ruter2010,Regensburger2012,Peng2014,Hodaei2014}, electrical circuits~\cite{Schindler2011, Chitsazi2017,Wang2020}, mechanical oscillators~\cite{Bender2013}, acoustics~\cite{Zhu2014}, and viscous fluids~\cite{Humire2019}. Post-selection over no-quantum-jump trajectories has further enabled their realizations in minimal quantum systems such as an NV center~\cite{Wu2019}, a superconducting qubit~\cite{Naghiloo2019}, ultracold atoms~\cite{Li2019}, or correlated photons~\cite{Klauck2019}. Concurrently,  open systems with anti parity-time ($\mathcal{APT}$) symmetry have emerged. A system has $\mathcal{APT}$-symmetry if its Hamiltonian $H$ {\it anticommutes} or its Liouvillian $L\equiv-iH$ commutes with the $\mathcal{PT}$ operator.  As a result, the eigenvalues $(\lambda_m,\lambda^*_m)$ of the Liouvillian $L$ are purely real or complex conjugates~\cite{Ruzicka2021}. Thus, while the symmetry-breaking transition in a $\mathcal{PT}$-symmetric system is marked by the emergence of amplifying and decaying eigenmode pairs, modes of an $\mathcal{APT}$-symmetric system amplify or decay independent of each other. $\mathcal{APT}$-symmetric systems have been realized in atomic vapor and cold atoms~\cite{Peng2016,Jiang2019}, active electrical circuits~\cite{Choi2018}, disks with thermal gradients~\cite{Li2019antiPT}, and diverse optical setups~\cite{Zhang2019,Fan2020,Zhang2020synthetic,Bergman2020}. 

Markovianity is a key feature of all deeply investigated $\mathcal{PT}$-symmetric, $\mathcal{APT}$-symmetric, or non-Hermitian systems. The first-order differential equation governing the state of such a system, $i\partial_t|\psi(t)\rangle=H_\mathrm{eff}[t;\psi(t)]|\psi(t)\rangle$, ensures that the rate of change of $|\psi(t)\rangle$ depends only on the system's properties at time $t$ and not on its history. This includes cases with a nonlinearity where the effective Hamiltonian $H_\mathrm{eff}$ depends on $\psi(t)$~\cite{Konotop2016}. Markovian (or memoryless) nature of such effective non-Hermitian dynamics is considered inviolate, although no fundamental principles prohibit it. 

Here, we propose a non-Markovian $\mathcal{APT}$-symmetric system where a single time-delay $\tau$ encodes the memory, and experimentally demonstrate its consequences in a system of two semiconductor lasers with bidirectional, time-delayed feedback~\cite{Chembo2019,Soriano2013}. A transcendental equation with infinitely many eigenvalues, that results from the non-local-in-time nature of a delay-differential equation, distinguishes our model from its Markovian counterpart with a quadratic eigenvalue equation.

We analytically obtain predictions for the key features of steady-state intensity profiles as a function of non-Markovianity, i.e. the time delay. These predictions coincide exceptionally well with results from numerical simulations of time-delayed, nonlinear, modified Lang-Kobayashi (LK) equations for the two electric fields $E_{1,2}(t)$ and the corresponding excess carrier inversions $N_{1,2}(t)$ in the two lasers~\cite{Soriano2013,Lang1980,Mulet2002}, and thereby validate our minimal, non-Markovian, $\mathcal{APT}$-symmetric model. Experimental observations of the steady-state laser intensities $I_{1,2}$ as a function of bidirectional feedback strength $\kappa$, individual laser frequencies $\omega_{1,2}$, and the time delay $\tau$ match our (analytical/numerical) predictions qualitatively, but not quantitatively. These robust signatures, clearly observed in experiments with off-the-shelf equipment and no custom fabrications, indicate that non-Markovianity (or time delay) opens up a new dimension for non-Hermitian systems. 

\begin{figure*}
\centering
\includegraphics[width=\textwidth]{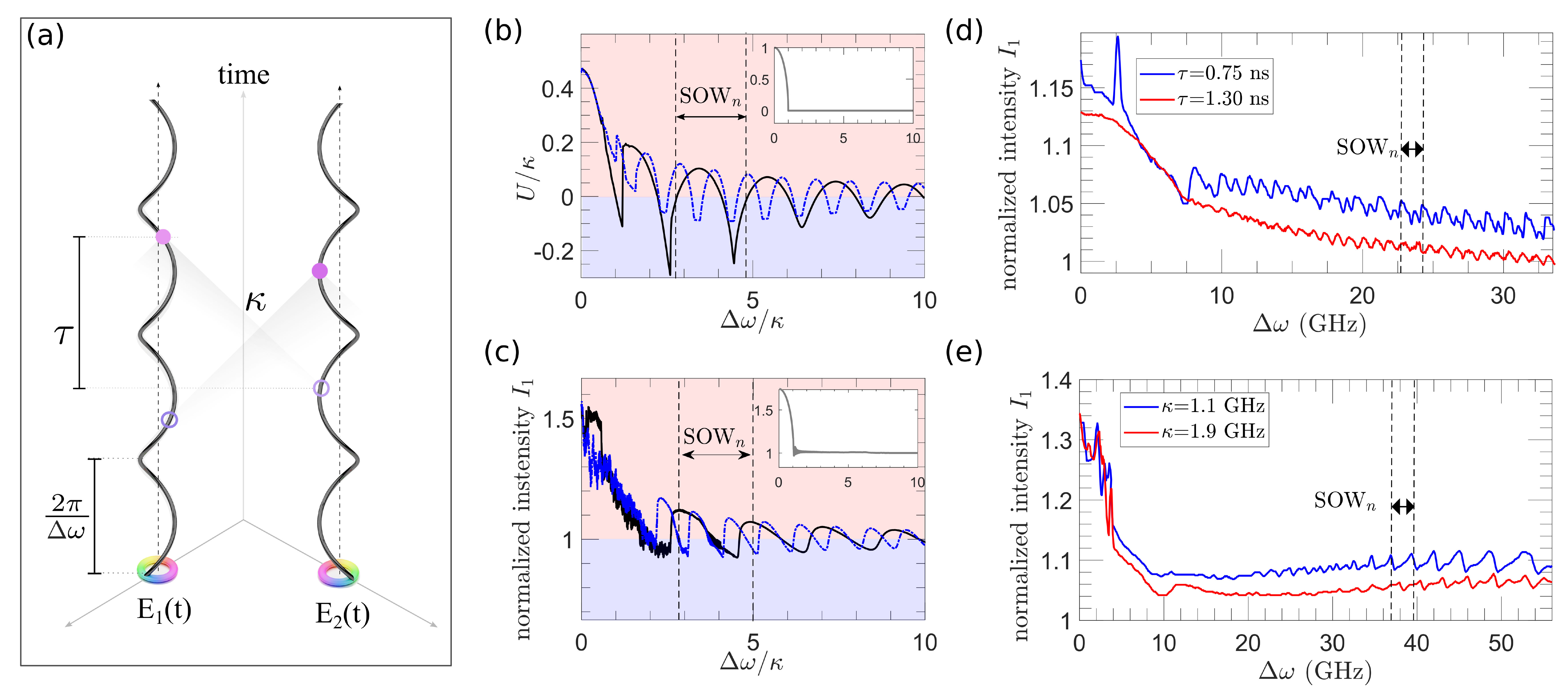}
\caption{Non-Markovian $\mathcal{APT}$-symmetric system. (a) Two modes $E_{1,2}(t)$ evolve with opposite phases $\pm\Delta\omega t$ in frame rotating with frequency $\omega_0$. Due to finite speed of light, each mode at time $t$ (filled circles) couples to the other at an earlier time $t-\tau$ (open circles). This non-Markovian coupling $\kappa$ is shown along the (shaded) past light-cones. This model, described by Eq.(\ref{eq:model}), is experimentally realized with two semiconductor lasers with bidirectional, time-delayed feedback; see Fig.~\ref{fig:schematic}. (b) Amplification rate $U_\tau$ shows sideband oscillations with a constant width (SOW) (solid black traces).  Results for $U_\mathrm{exp}$ show that the SOW is halved (dashed blue traces).  $U>0$ region (pink) denotes amplifying modes, while $U<0$ region (violet) denotes decaying modes. Inset: in the Markovian limit $\tau=0$, the $\mathcal{APT}$ transition from $U>0$ to $U=0$ occurs at $\Delta\omega=\kappa$. (c) Steady state intensity $I_{1}(\Delta\omega)$ obtained from four, coupled, nonlinear rate equations shows sideband oscillations whose constant width is halved when $\omega_1$ is varied ($L_\mathrm{exp}$; dashed blue traces) instead of varying $\Delta\omega$ while keeping $\omega_0$ constant ($L_\tau$; solid black traces). Despite obvious similarities, explicit mapping from $U(\Delta\omega)$ to the steady-state $I_{1,2}(\Delta\omega)$ is unknown. Inset: At $\tau=0$, a central dome at small detuning changes into a flat intensity profile for $\Delta\omega\geq\kappa$. (d) Exemplary traces of experimentally measured intensity $I_{1}(\Delta\omega)$ obtained by sweeping $\omega_1$ at $\tau$=0.75 ns (blue) and $\tau$=1.3 ns (red) show that observed SOW is reduced with increasing $\tau$. Their features are consistent with our model and full laser dynamics simulations.  (e) Exemplary traces of intensity $I_1(\Delta\omega)$ obtained at $\kappa$=1.1 GHz (blue) and $\kappa$=1.9 GHz (red) show that the observed SOW is insensitive to the coupling $\kappa$. The central dome in (b)-(e) at small $\Delta\omega$ is present in the Markovian limit ($\tau=0$) and signals the standard $\mathcal{APT}$-transition. We analytically determine the behavior of the key non-Markovian signature $\mathrm{SOW}_n(\kappa,\tau)$ for $L_\tau$ and $L_\mathrm{exp}$.
}
\label{fig:fig1}
\end{figure*}


\noindent{\bf Time-delayed $\mathcal{APT}$-symmetric model.} For a system of two modes $E_{1,2}(t)$ with free-running frequencies $\omega_{1,2}=\omega_0\pm\Delta\omega$ and time-delayed coupling (Fig.~\ref{fig:fig1}a), in a frame rotating at the center frequency $\omega_0$, the dynamics are described by 
\begin{equation}
\label{eq:model}
\begin{array}{r@{}l}
\partial_t E_1 &{}= i\Delta\omega E_1(t)+\kappa e^{-i\omega_0\tau} E_2(t-\tau),\\
\partial_t E_2 &{}=-i\Delta\omega E_2(t) +\kappa e^{-i\omega_0\tau}E_1(t-\tau). 
\end{array}
\end{equation}
This model emerges from the microscopic rate equations (Supplementary Materials) for full dynamics of two, nominally identical, bidirectionally delay-coupled semiconductor lasers~\cite{Soriano2013,Lang1980,Mulet2002} operating in the single-mode regime with with vanishing excess carrier densities $N_{1,2}$. At zero delay, Eq.(\ref{eq:model}) reduces to $\partial_t\vec{E}(t)=L\vec{E}(t)$ where $\vec{E}(t)=[E_1(t),E_2(t)]^T$, the Liouvillian is given by $L(\Delta\omega,\kappa)=i\Delta\omega\sigma_z+\kappa\sigma_x$, and $\sigma_z,\sigma_x$ are standard Pauli matrices. It describes a Markovian $\mathcal{APT}$-symmetric system where $\mathcal{P}=\sigma_x$ and $\mathcal{T}$ is complex conjugation. When the detuning $\Delta\omega$ is increased, the eigenvalues $\lambda_{\pm}=\pm\sqrt{\kappa^2-\Delta\omega^2}$ of the Liouvillian change from real to complex conjugates, and the amplifying/decaying modes change into oscillatory ones with constant intensity. In reality, the nonlinearity of the gain medium saturates the exponentially amplifying mode intensities $I_{1,2}(t)=|E_{1,2}(t)|^2$ into steady-state values that monotonically decrease with $\Delta\omega$ and become constant when $\Delta\omega\geq\kappa$ (Fig.~\ref{fig:fig1}b,c insets). Although the experimentally accessible steady-state intensities $I_{1,2}$ scale monotonically with the analytically derived amplification rate $|\mathrm{Re}\lambda_\pm|$, their  functional dependence is unknown. 

When $\tau>0$, Eq.(\ref{eq:model}) becomes $\partial_t\vec{E}=\mathcal{L}\vec{E}$ where the non-local Liouvillian contains the time-delay operator, 
\begin{equation}
\label{eq:Ltau}
\mathcal{L}(\Delta\omega,\kappa,\omega_0,\tau)= i\Delta\omega\sigma_z+\kappa e^{-i\omega_0\tau} e^{-\tau\partial_t}\sigma_x. 
\end{equation}
If the two mode-frequencies are swept antisymmetrically while maintaing $\omega_0$ at $e^{i\omega_0\tau}=\pm 1$, the Liouvillian commutes with $\mathcal{PT}$ where the $\mathcal{T}$-operator also takes $\tau$ to $-\tau$. Then this non-Markovian system has $\mathcal{APT}$ symmetry. We will denote this Liouvillian as $L_\tau\equiv i\Delta\omega\sigma_z+\kappa e^{-\tau\partial_t}\sigma_x$. The characteristic equation for the eigenmodes $\vec{E}(t)=\exp(\lambda t)\vec{E}(0)$ of $L_\tau$ is given by 
\begin{equation}
\label{eq:char1}
\lambda^2+\Delta\omega^2-\kappa^2 e^{-2\lambda\tau}=0.
\end{equation}
This transcendental equation has infinitely many eigenvalue pairs $(\lambda_m,\lambda^*_m)$.  Experimentally it is easier to sweep $\omega_1$ while keeping $\omega_2$ constant, which changes $\omega_0$ and $\Delta\omega$ in a correlated manner. We call the corresponding Liouvillian $L_\mathrm{exp}$, and it is given by 
\begin{equation}
\label{eq:Lexp}
L_{\mathrm{exp}}=i\Delta\omega\sigma_z + \kappa e^{-i(\omega_1-\Delta\omega)\tau} e^{-\tau\partial_t}\sigma_x.
\end{equation}
Since this Liouvillain does not commute with the $\mathcal{PT}$ operator, its eigenvalues $\lambda_m$ are neither complex-conjugate pairs nor symmetric in $\Delta\omega\leftrightarrow-\Delta\omega$. The long-time, steady-state dynamics of the system are determined by the effective amplification rate $U\equiv\max\Re\lambda_m$. A positive $U$ means that there is an amplifying mode, while $U<0$ means all modes are below the lasing threshold. 

Figure~\ref{fig:fig1}b shows the numerically obtained $U_\tau$ from $L_\tau$ (solid black line) and $U_\mathrm{exp}$ from $L_\mathrm{exp}$ (dashed blue line) as a function of dimensionless detuning $\Delta\omega/\kappa$ when the time delay is $\kappa\tau$=2. Apart from the central dome present in the $\tau$=0 limit (Fig.~\ref{fig:fig1}b inset), both show time-delay induced sideband oscillations whose width, SOW, is constant at large $\Delta\omega$. The SOW for $L_\tau$ is twice as large as it is for $L_\mathrm{exp}$. For these two system configurations, we obtain the steady-state intensities by solving four modified  LK equations (Supplementary Material). The results for the intensity of the first laser,  normalized to its large $\Delta\omega/\kappa$ value (Fig.~\ref{fig:fig1}c), also show sidebands with an SOW that is twice as large for the $\mathcal{APT}$-symmetric model (solid black line) as it is for the experimental setup (dashed blue line); these sidebands are absent in the Markovian limit (Fig.~\ref{fig:fig1}c inset). 

The striking similarity of results in (b)-(c), occurring over a wide range of time delays and feedback~\cite{AWthesis2021}, indicates that our minimal model captures key signatures of non-Markovianity that emerge from four, delay-coupled, nonlinear rate equations. Figure~\ref{fig:fig1}d shows exemplary experimental traces for normalized intensity $I_1(\Delta\omega)$ at $\tau$=1.3 ns (red) and $\tau$=0.75 ns (blue), at $\kappa$=3.1 GHz. Clear sidebands are visible with an SOW that decreases with increasing delay-time. Conversely, experimental traces in Fig.~\ref{fig:fig1}e for normalized $I_1(\Delta\omega)$ at $\kappa$=1.9  GHz (red) and $\kappa$=1.1 GHz (blue) at a fixed time delay $\tau$=0.75 ns show that the SOW is largely insensitive to the coupling. Experimental data over a wide range of $\kappa$ and $\tau$~\cite{AWthesis2021} indicate that, while the central dome width $\Delta\omega_c$ and sideband oscillation amplitudes depend on both, the SOW is solely determined by the time delay. 

\noindent{\bf SOW theory and experimental results.} Emergence of constant-width oscillations in the steady-state intensity is the key signature of non-Markovianity on an $\mathcal{APT}$-symmetric system. To analytically determine SOW$(\kappa,\tau)$, we investigate the flow of eigenvalues $\lambda=u+iv$ of $L_\tau$. It is best understood via common zeros 
of two real functions comprising Eq.(\ref{eq:char1}), 
\begin{align}
\label{eq:F} 
&F(u,v)=u^2-v^2+\Delta\omega^2-\kappa^2e^{-2u\tau}\cos(2v\tau),\\
\label{eq:G}
&G(u,v)=2uv+\kappa^2e^{-2u\tau}\sin(2v\tau).
\end{align}

\begin{figure}
\centering
\includegraphics[width=\columnwidth]{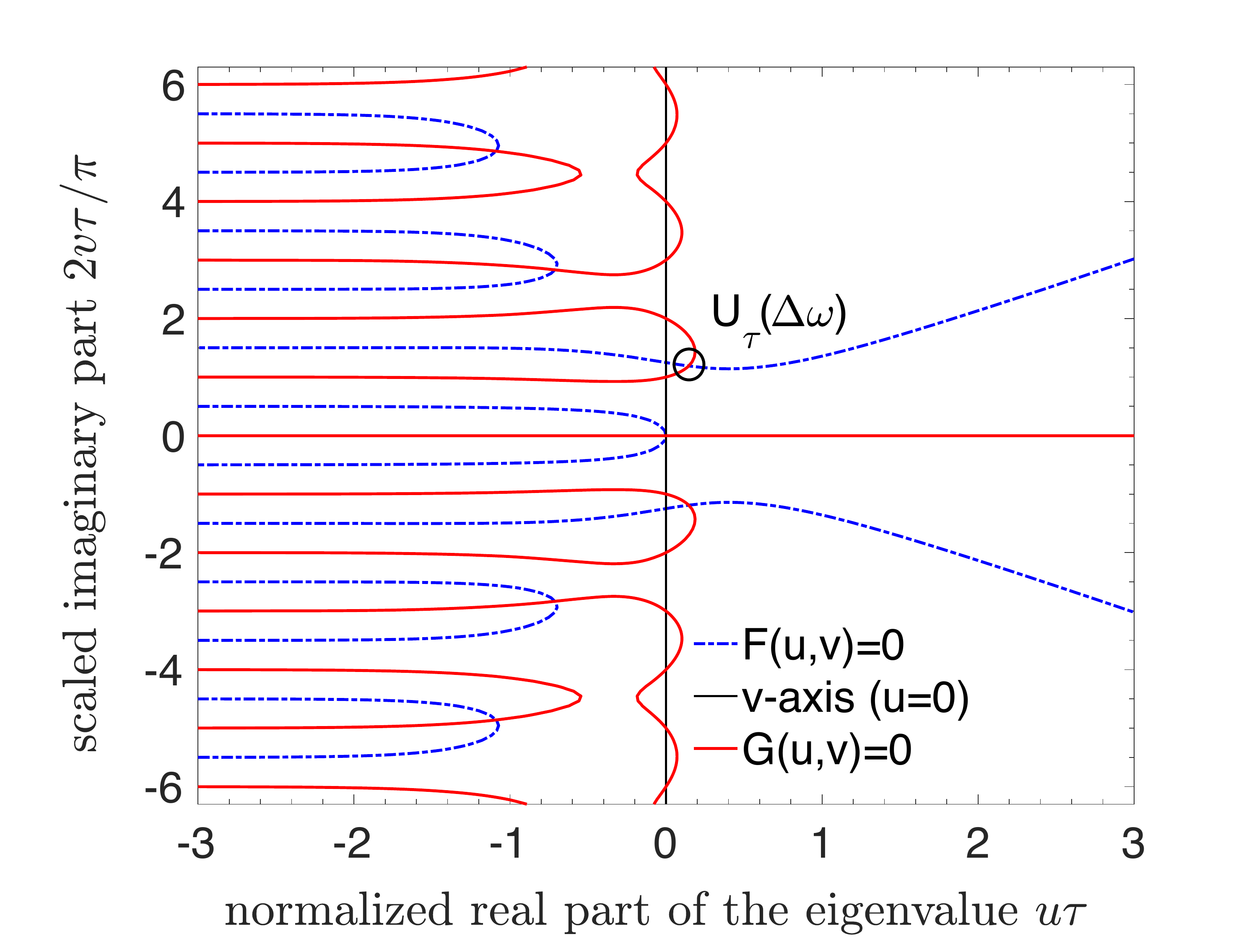}
\caption{Eigenvalues $\lambda=u+iv$ of Liouvillian $L_\tau$ occur at the intersections of $F(u,v)=0$ and $G(u,v)=0$ contours, shown here for $\kappa\tau=1.5$ and $\Delta\omega/\kappa=1$. Properties of $G(u,v)=0$ contours and their intersections with the two axes are analytically determined by the Lambert $W$ function~\cite{Corless1996,Joglekar2017}. At small detuning, the hyperbolic $F(u,v)=0$ contour always intersects the $u>0$ axis and gives the central dome that survives in the Markovian limit. At large $\Delta\omega$, intersections of the $G=0$ and $F=0$ contours on the vertical axis ($u=0$) give an infinite sequence of $U_\tau(\Delta\omega)>0\leftrightarrow U_\tau(\Delta\omega)<0$ transitions that manifest as sideband oscillations seen in Fig.~\ref{fig:fig1}b-e.}
\label{fig:fig2}
\end{figure}

The $U_\tau>0\leftrightarrow U_\tau<0$ transitions that lead to the sidebands occur when $G=0$ and $F=0$ contours intersect in the vicinity of the vertical $v$-axis (Fig.~\ref{fig:fig2}). By determining the detunings $\Delta\omega_n$ at which they occur the $\mathrm{SOW}_n\equiv(\Delta\omega_{n+2}-\Delta\omega_n)$ is obtained (Fig.~\ref{fig:fig1}b). Since the contours of $G=0$ are independent of $\Delta\omega$, we characterize them first (Fig.~\ref{fig:fig2} solid red lines). When $-u\tau\gg 1$, due to the divergent exponential factor, lines $v_m=m\pi/2\tau$ ($m\in\mathbb{Z}$), parallel to the $u$-axis, are solutions of $G=0$. These points, i.e. $0+iv_m$, also satisfy $G=0$ along at $v$-axis, as does the entire $u$-axis ($v=0$). In addition to these simple zeros, $\partial_vG(u,0)=0$ determines the double zeros along the $u$-axis. They are given by values of $z=2u\tau$ that satisfy the equation $ze^z=-2(\kappa\tau)^2$. Thus, for $\kappa\tau<1/\sqrt{2e}\approx 0.43$, there are two negative solutions $u_{0,1}=W_{0,1}(-2\kappa^2\tau^2)/2\tau$ where $W_m(x)$ is the Lambert $W$ function~\cite{Corless1996,Joglekar2017}. We note that $u_1$ is the intersection of the $v_{\pm 1}=\pm\pi/2\tau$ branches with the $u$-axis, while $u_0$ is the intersection of the deformation of the $v$-axis, which is a solution of $G=0$ at zero delay. 

Next, let us consider the evolution of $F=0$ contours when $\Delta\omega$ is varied at a fixed $\tau$ (Fig.~\ref{fig:fig2} dashed blue lines). When $-u\tau\gg 1$, parallel lines at $v_{m'}=(2m'+1)\pi/4\tau$ ($m'\in\mathbb{Z}$) are solutions of $F=0$. At $u\tau\gg 1$, they are given by hyperbolas $v=\pm\sqrt{u^2+\Delta\omega^2}$. For small $\Delta\omega$, the $F=0$ contour intersects with the positive $u$-axis at $z'=u\tau$ that satisfies $z'e^{z'}=\kappa\tau[1-(\Delta\omega e^{z'}/\kappa)^2]^{1/2 }$. The solution reduces from $u_+=W_0(\kappa\tau)/\tau$ when $\Delta\omega=0$ to zero as $\Delta\omega\rightarrow\Delta\omega_c\sim\kappa$. It corresponds to the amplifying mode underlying the central dome that persists in the Markovian limit (Fig.~\ref{fig:fig1}b-e).


\begin{figure}
\centering
\includegraphics[width=\columnwidth]{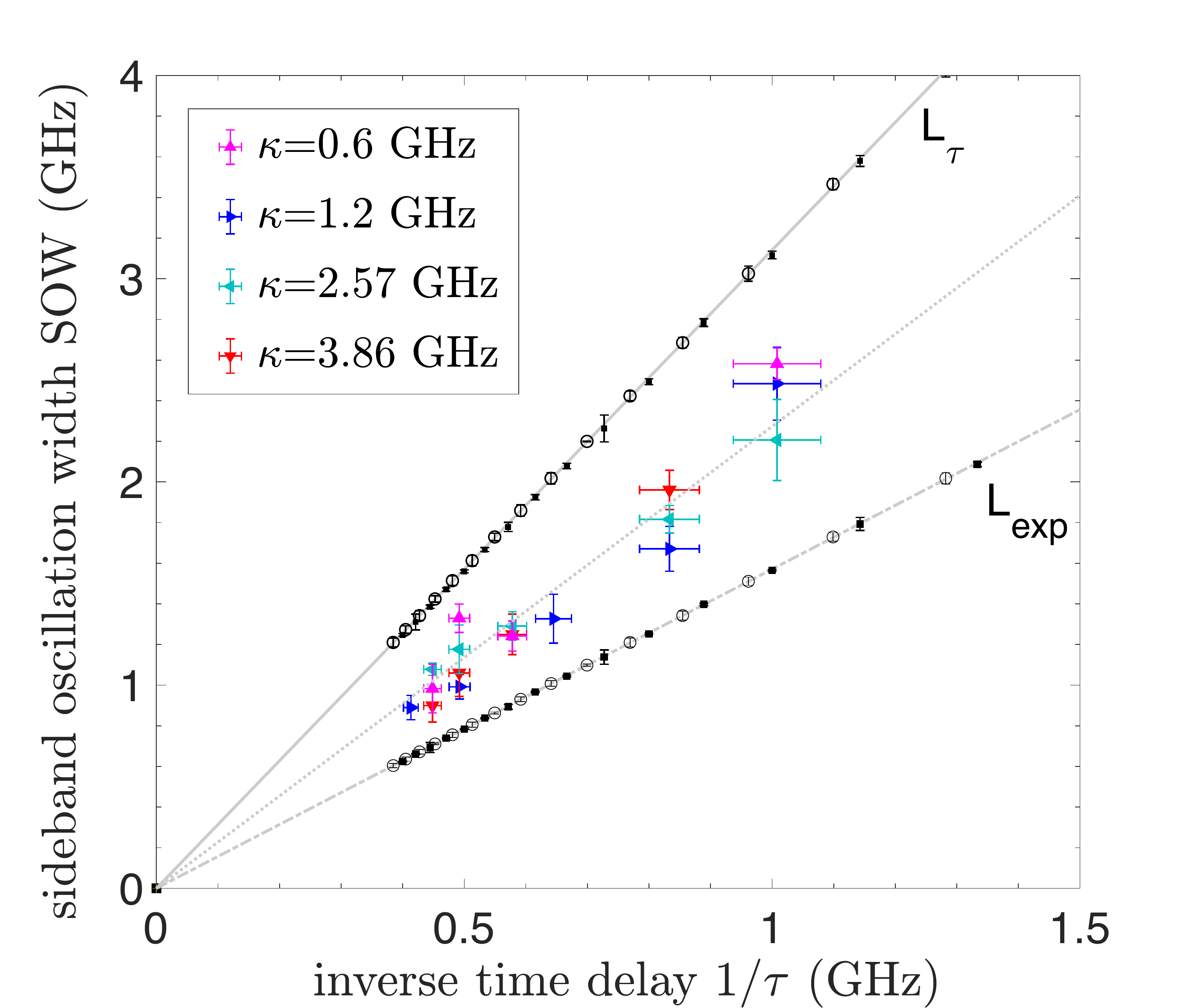}
\caption{Time-delay induced transitions. Eigenvalue analysis predicts $\mathrm{SOW}\propto 1/\tau$ with $\kappa$-independent prefactor of $\pi$ for $L_\tau$ (gray solid curve) and $\pi/2$ for $L_\mathrm{exp}$ (gray dot-dashed curve). SOWs extracted from steady-state intensity sidebands obtained from the full LK simulations match the eigenvalue predictions exceptionally well; error bars, obtained from FWHM of the Fourier transform, are smaller than symbols when not shown. $\kappa$-independence of the prefactor in full LK simulations is clear ( squares: $\kappa$=2 GHz; circles: $\kappa$=0.4 GHz). $\text{SOW}(\tau)$, obtained from experimental data for $\kappa$ that varies by a factor of six, clearly show a $1/\tau$ behavior. Vertical error-bars are FWHM of the sideband Fourier transform; horizontal error-bars in time-delay estimate are from a fixed uncertainty $\Delta l$=1 cm in the optical path length.}
\label{fig:fig3}
\end{figure}

At larger $\Delta\omega$, the $F=0$ contours intersect the $v$-axis, at two, mirror-symmetric intersections $(0,\pm\bar{v})$. As the zeros of $G$ are at $n\pi/2\tau$, the most dominant eigenvalue $\lambda=0^{\pm}+i\bar{v}$ changes from positive to negative when $\bar{v}$ traverses ``even $n$'' branches of $G=0$ contours. When $\bar{v}=v_n$, this leads to $\Delta\omega_n=\sqrt{v_n^2+\kappa^2}$. Therefore, we predict that
\begin{equation}
\label{eq:SOW}
\mathrm{SOW}_n(\kappa,\tau)=\frac{\pi}{\tau}-\frac{\kappa^2\tau}{\pi n(n+2)}\xrightarrow[n\gg1]{} \frac{\pi}{\tau}.
\end{equation}
With a similar analysis for eigenvalues of $L_\mathrm{exp}$, Eq.(\ref{eq:Lexp}), we find that the SOW is reduced by a factor of two, i.e. $\mathrm{SOW}(\kappa,\tau)=\pi/2\tau$, because $\Delta\omega$ generated by varying $\omega_1$ with a fixed $\omega_2$ is half of what is generated when $\omega_1$ and $\omega_2$ are varied antisymmetrically. Figure~\ref{fig:fig3} shows these predictions for $L_\tau$ (solid gray) and $L_\mathrm{exp}$ (dot-dashed gray) as lines with slopes $\pi$ and $\pi/2$ respectively. 

To validate the eigenvalue-analysis predictions, we obtain the SOWs from full laser-dynamics simulations for the two cases~\cite{Lang1980,Mulet2002}. Steady state intensities $I_{1,2}(\Delta\omega|\kappa,\tau)$ are obtained over a range of $\Delta\omega$ such that $\gtrsim 20$ sidebands are present away from the central dome. For a given $\kappa$ and $\tau$, SOW is obtained by Fourier transform of the sideband data; the error-bars indicate full width at half maximum (FWHM) of the single peak that is present in the Fourier transform. The results from such analysis carried out for delay times $\tau$ ranging from 0.6 ns to 2.5 ns are plotted in Fig.~\ref{fig:fig3}. They are obtained for $\kappa$=0.4 GHz (open circles) and $\kappa$=2 GHz (filled squares), and yet SOWs derived from the full LK simulations do not depend on $\kappa$. Their striking agreement with the analytical predictions shows that our minimal models, defined by $L_\tau$ and $L_\mathrm{exp}$, capture the key consequences of introducing non-Markovianity in non-Hermitian ($\mathcal{APT}$-symmetric) systems. 


We obtain experimental laser intensity profiles $I_{1,2}(\Delta\omega)$ by changing the temperature and consequently the frequency $\omega_1$ of the first laser, and normalize each 
by the minimum recorded intensity at large detuning. Since the amplitude of sideband oscillations is small, we average $\sim20$ oscillations away from the central dome to obtain the SOW. Figure~\ref{fig:fig3} shows that the experimentally obtained $\mathrm{SOW}(\tau)$ varies inversely with delay time $\tau$, and essentially remains unchanged when the feedback strength $\kappa$ is varied over a factor of six. The slope of the experimental data for $\mathrm{SOW}$ vs. $1/\tau$ best-fit line (dotted gray) is halfway between the predictions for the $\mathcal{APT}$-symmetric $L_\tau$ model and non-Hermitian $L_\mathrm{exp}$ model, but two key features of Eq.(\ref{eq:SOW})---namely, $1/\tau$ variation and vanishing $\kappa$ dependence---are robustly retained. 


\noindent{\bf Discussion.} Delay differential equations model systems from engineering, physics, chemistry, biology, and epidemiology~\cite{Kyrychko2010,Otto2019,Marc1996,Glass2021,Luce2020}, and exhibit synchronization, bifurcation, and chaos~\cite{Erneux2017,Otto2019}. They have long been used for classical random-number generation and control~\cite{Wang19,Ma20,Just1997,Pyragas2006}. We have shown that non-Markovianity via time delay adds a novel dimension to the verdant field of non-Hermitian open systems. Our choice of $\mathcal{APT}$-symmetric model is motivated by a standard setup of two, bidirectionally coupled semiconductor lasers. We have mapped the complex, nonlinear system into simple, analytically tractable non-Markovian models. Their multifarious dynamics contain robust signatures of transitions that occur solely due to the non-Markovianity. We find that predictions from the minimal models quantitatively capture those from the full laser dynamics model. Their variance from the experimental data is likely due to the failure of the single-mode approximation or the weak coupling approximation, and possible variation of the second-laser frequency when the frequency of the first laser is varied. 

We have considered a system with $\mathcal{APT}$-symmetry. Its Wick-rotated counterpart, i.e. a $\mathcal{PT}$- symmetric system with time delay, can naturally arise in electrical oscillator circuits and classical wave systems. In the quantum domain, $\mathcal{PT}$-symmetric systems have been realized through post-selection on a minimal quantum system coupled to an environment~\cite{Wu2019,Naghiloo2019,Klauck2019,Li2019}. Coherent feedback with time-delay has been proposed as a control mechanism for precisely such open quantum systems~\cite{Zoller1996,ask2021}. Investigation of non-Markovianity induced phenomena in such systems remains an open question.


\noindent{\bf Acknowledgments.} We thank Kaustubh Agarwal for help with Figure 1. The views and opinions expressed in this paper are those of the authors and do not reflect the official policy or position of the U.S. Air Force, Department of Defense, or the U.S. Government. 

\bibliographystyle{apsrev4-2}

%


\clearpage
\pagebreak
\widetext
\begin{center}
	\textbf{\large Supplementary Material}\\ \vspace{9pt}
\end{center}

\setcounter{equation}{0}
\setcounter{figure}{0}
\setcounter{table}{0}
\setcounter{page}{1}
\makeatletter
\renewcommand{\theequation}{A\arabic{equation}}
\renewcommand{\thefigure}{S\arabic{figure}}
\makeatother


\section{Numerical Model}
\label{sec:suppa}
Our system of two, coupled semiconductor lasers is described by a modified version of the well-known Lang-Kobayashi model~\cite{Soriano2013,Lang1980,Mulet2002} for a solitary, single-mode laser with weak, time-delayed feedback. Two identical, single-mode lasers are operating at free-running frequencies $\omega_{1,2}=\omega_0\pm\Delta\omega$.  The slowly varying envelopes $E_{1,2}(t)$ of the electric fields inside the rectangular laser cavities are defined in a reference frame rotating at the average $\omega_0$ of the two optical frequencies. The rate equations describing the complex electric fields  and excess carrier densities $N_{1,2}(t)$ can be written as follows~\cite{Lang1980,Mulet2002}						
\begin{eqnarray}
\label{eq:e1}
\frac{dE_1}{dt}&=& \frac{1}{2}(1+i\alpha)GN_1(t)E_1(t) + {i\Delta\omega}E_{1}(t)+\frac{\mathcal{K}}{\tau_\text{in}}e^{-i\omega_0\tau}E_{2}(t-\tau),\\
\label{eq:e2}
\frac{dE_{2}}{dt}&=& \frac{1}{2}(1+i\alpha)GN_2(t)E_{2}(t) - {i\Delta\omega}E_{2}(t) + \frac{\mathcal{K}}{\tau_\text{in}}e^{-i\omega_0\tau}E_{1}(t-\tau),\\
\label{eq:n1}
\frac{dN_{1}}{dt}&=& J_{1} -\frac{N_\text{th}}{\tau_s}- \frac{N_{1}(t)}{\tau_s} - \left[\frac{1}{\tau_p}+G N_1(t)\right]|E_{1}(t)|^2,\\
\label{eq:n2}
\frac{dN_{2}}{dt}&=& J_{1} -\frac{N_\text{th}}{\tau_s}- \frac{N_{2}(t)}{\tau_s} - \left[\frac{1}{\tau_p}+G N_2(t)\right]|E_{2}(t)|^2.
\end{eqnarray}
Here $\alpha$  is the linewidth enhancement factor, $G$ is the gain rate, $\mathcal{K}$ is the dimensionless  feedback strength, $\tau_\text{in}$ is the internal round-trip time for the laser cavity (< 1  ps), $\tau=l/c$  is the time delay, and $l$ is the free-space optical path length. In the excess-carrier-density equations,  $J_{1,2}$ are the injection current above threshold. $N_\text{th}$ is the steady-state inversion, $\tau_p$ is the photon lifetime (10 ps) and $\tau_s$ is the carrier lifetime (1 ns). This model has been used with great success to describe the nonlinear dynamical behavior of coupled semiconductor lasers~\cite{Soriano2013,Lang1980,Mulet2002}. 

We numerically solve the four coupled, nonlinear, time-delayed equations by using fourth-order Runge-Kutta method with 0.1 ps time-step increment. The seed solutions for the history $E_{1,2}(-\tau\leq t\leq 0)$ are obtained by solving the corresponding $\tau=0$ equations~\cite{AWthesis2021}. If the excess carrier densities for the steady-state solution are zero, $N_{1,2}=0$, it follows that the electric-field envelope equations, Eqs.(\ref{eq:e1})-(\ref{eq:e2}), reduce to Eq.(\ref{eq:model}) in the main text with $\kappa=\mathcal{K}/\tau_\text{in}$.  


\section{Experimental set up}
\label{sec:suppb}

\begin{figure}[h]
\includegraphics[width=\columnwidth]{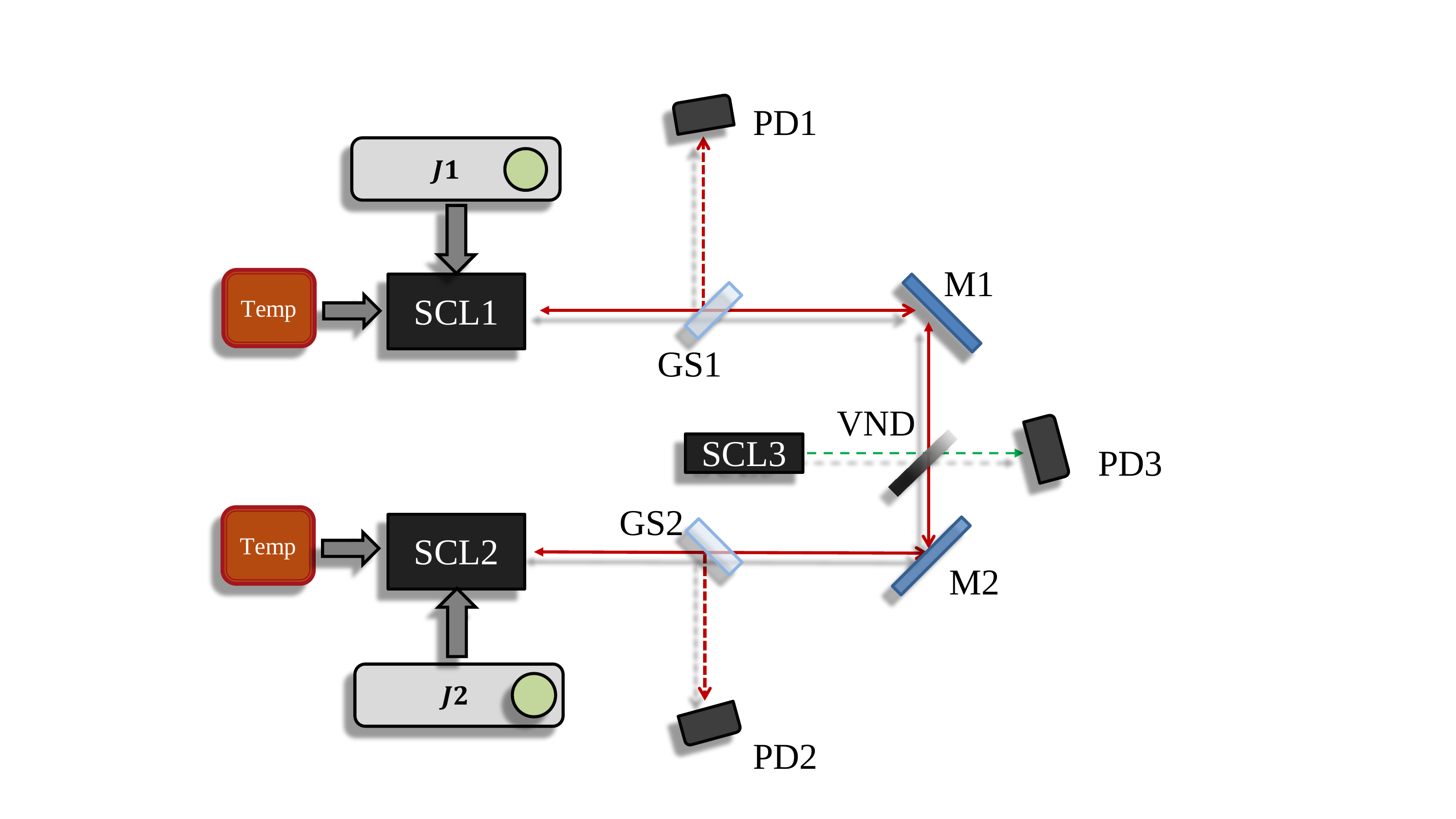}
\caption{Schematic of the experimental setup. Two nominally identical semiconductor lasers SCL1 and SCL2 are controlled by pump currents J1 and J2 respectively, and independent temperature controllers. The glass slide GS1 (GS2) reflects a small amount (8\%) of light to the photodiode PD1 (PD2) to measure steady-state laser intensities $I_{1,2}$. Mirrors M1 and M2, along with the variable neutral density (VND) filter provide bidirectional, time-delayed feedback. The transmission through the VND, and therefore the coupling $\kappa$, is determined by using another laser SLC3 along with a photodiode PD3.}
\label{fig:schematic}
\end{figure}

Our experimental system is shown in Fig.~\ref{fig:schematic}. It consists of two, identical single mode (HL7851G) semiconductor lasers (SCL1, SCL2), an external cavity consisting of two beam splitters (BS1 and BS2) which optically coupled the two lasers, and an external control of the coupling strength, $\kappa_\text{exp}=\mathcal{K}/\tau_\text{in}$ , via a variable neutral density filter (VND). 

The transmission through the VND is determined by an independent laser (SL3) and a photodiode (PD3) which allows us to calibrate the coupling strength as $\kappa_\text{exp}=\mathcal{K}/\tau_\text{in}=(r^{-1}-r)\zeta\tau_p/\tau_\text{in}$. Here $r<1$ is the reflectivity of the external laser facet, $\zeta^{2}$ is the fraction of optical power transmitted by all the optical elements, $\tau_\text{in}$ is the internal round-trip time, and $\tau_{p}$ is the photon lifetime.  Once the transmission through the VND is recorded, $\zeta ^{2}$ can be determined since all the other optical elements are fixed.  This model assumes that the fractional power is fully coupled into the active region of the semiconductor lasers.  However, due to the relative sizes of the beam profile (> 100  $\mu$m) and the active region (about 10 $\mu$m), only a portion of the power is coupled into the active region. Through literature reports and comparison of our experiments to LK simulations, we find that the ``effective coupling strength'' is reduced by a factor of ten, i.e. $\kappa_\text{LK}=\kappa_\text{exp}/10$.  

The experiment is designed such that the light coupling from the first laser into the second is equal to light coupling from the second into the first. A Faraday rotator is placed in the coupling beam path to eliminate self coupling. The glass slides GS1 and GS2 independently reflect a small portion of (8\%) of the intensity from SCL1 and SCL2 respectively to corresponding 1 GHz photodiodes (PD1 and PD2) in conjuction with a 1 GHz oscilloscope. The pump currents $J_{1,2}$ and temperatures $T_{1,2}$ of the lasers SCL1 and SCL2 are stabilized to 10 $\mu$A and 0.01 C respectively. 

After bidirectionally coupling the two lasers, the temperature of SCL1 is slowly scanned  (< 10 Hz) and the steady-state intensities $I_{1.2}$ of the two lasers are monitored.  The  key parameters $\kappa$ and $\Delta\omega$ can be varied via the VND and the temperature of SCL2, respectively. For our room-temperature lasers, the frequencies $\omega_{1,2}$ are proportional to the temperature of the relevant  active region. For a small temperature range (< 4 C), we use a linear approximation for the laser frequency $\omega$ and intensity $I$, 

\begin{eqnarray}
\label{eq:w1}
\omega(T)&=&\omega(T_0)-A_{T}(T-T_0),\\
\label{eq:i1}
I(T)&=&I(T_0)+B_{T}(T-T_0),
\end{eqnarray}
where the coefficients $A_{T}$=20 GHz/C and  $B_{T}$=0.15 mW/C are experimentally determined for our setup.  $A_{T}$ is experimentally determined by scanning the temperature of a laser while monitoring the transmitted intensity through a fixed 2 GHz free spectral range Fabry-Perot etalon.  When $n$ peaks are observed through the etalon, we obtain $A_{T}= 2n/\Delta T$ GHz/C where ${\Delta}T$ is the range of temperature scanned. To obtain $B_{T}$, the scanned temperature and the emitted laser light intensity are recorded and fit to Eq.(\ref{eq:i1}). 

For detuning beyond the central dome, the analytical models ($L_\tau$ and $L_\text{exp}$) and simulations of the full laser model, Eqs.(\ref{eq:e1})-(\ref{eq:n2}), show that time-delay causes oscillations in the steady-state intensities as a function of $\Delta\omega$. We assign these oscillations to those in the sign of $U_\tau$ or $U_\text{exp}$.  To test this hypothesis, we operate the two lasers at constant injection currents barely (3\%) above their stand-alone lasing thresholds.  This guarantees that they remain above threshold when the temperature is scanned. The optical spectrum of the uncoupled (free-running) lasers is independently measured by a scanning an optical spectrum analyzer.  The temperature to one laser was scanned and recorded.  Using Eq. (8) along with temperature and wavelength measurements, we calculated the detuning, $\Delta\omega$.  The temperature, coupling strength and the two SCL intensities were simultaneously recorded resulting in the reported intensity profiles of Fig. 1.  The photodiodes have a large load resistor that decreases  the bandwidth to < 1 GHz.  This bandwidth, along with the scan rate of the oscilloscope, leads to intensities $I_{k}$ that are averaged over a time-window of 1 ns. 


\end{document}